\documentclass[epj]{svjour}

\usepackage{graphicx}
\usepackage{dcolumn}
\usepackage{amsmath}
\usepackage{mathrsfs}
\usepackage{amssymb}
\usepackage{url}
\begin{document}
\title{Nuclear level density and thermal properties of $^{115}$Sn from neutron evaporation}
\author{Pratap Roy\inst{1,2}
\thanks{\emph{Email for correspondence:} roypratap@vecc.gov.in, pratapproy@gmail.com}
\and K. Banerjee\inst{1,2}
\and T. K. Rana\inst{1}
\and S. Kundu\inst{1}
\and Deepak Pandit\inst{1,2}
\and N. Quang Hung\inst{3,4}
\and T. K. Ghosh\inst{1,2}
\and S. Mukhopadhyay\inst{1,2}
\and D. Mondal\inst{1} 
\and G. Mukherjee\inst{1,2} 
\and S. Manna\inst{1,2}
\and A. Sen\inst{1,2} 
\and S. Pal\inst{1,2}
\and R. Pandey\inst{1}
\and D. Paul\inst{1,2} 
\and K. Atreya\inst{1,2} 
\and C. Bhattacharya\inst{1,2}
}
\institute{Variable Energy Cyclotron Centre, 1/AF, Bidhan Nagar, Kolkata - 700 064, India \and Homi Bhabha National Institute, Training School Complex, Anushakti Nagar, Mumbai - 400 094, India, \and Institute of Fundamental and Applied Sciences, Duy Tan University, Ho Chi Minh City 700000, Vietnam, \and Faculty of Natural Sciences, Duy Tan University, Danang city 550000, Vietnam}
%
\abstract{
The nuclear level density of $^{115}$Sn has been measured in an excitation energy range of $\sim $2~-~9 MeV using the experimental neutron evaporation spectra from the $^{115}$In($p,n$)$^{115}$Sn reaction. The experimental level densities were compared with the microscopic Hartree-Fock BCS (HFBCS), Hartree-Fock-Bogoliubov plus combinatorial (HFB+C), and an exact pairing plus independent particle model (EP+IPM) calculations. It is observed that the EP+IPM provides the most accurate description of the experimental data. The thermal properties (entropy and temperature) of $^{115}$Sn have been investigated from the measured level densities. The experimental temperature profile as well as the calculated heat capacity show distinct signatures of a transition from the strongly-paired nucleonic phase to the weakly paired one in this nucleus.
%
} 
\maketitle
%
\section{\label{sec1:intro} Introduction}
Nuclear level density (NLD), is one of the most critical inputs of the statistical Hauser-Feshbach (HF) calculation~\cite{Hauser} of compound nuclear reactions. NLDs are observed to have a strong impact on the calculated neutron-, and proton-capture rates relevant for the $r$-, and $p$-processes of heavy-element nucleosynthesis~\cite{Cecilie,Arnould}. Besides, a precise determination of NLD is critical for many practical applications in different areas such as the fusion-fission cross-section calculations for reactor simulations, modeling of the spallation reactions for the development of spallation neutron source and accelerator-driven sub-critical systems (ADSS)~\cite{Bowman}, and production yield estimation of radioisotopes for medical application~\cite{Mirz}. \\
Experimental NLD also acts as a testing ground for different nuclear structure models as well as provide crucial information on the thermodynamic properties of atomic nuclei~\cite{Melby,Giacoppo,Melby2,Agvaan,Toft,Syed,Gutt,Bala,Schil,Kaneko,Chankova}. Recent investigations on the thermal properties of several nuclei have revealed interesting features like the existence of peak-like structures in the microcanonical temperature~\cite{Melby,Giacoppo,Melby2,Agvaan,Toft}, S-shaped heat capacity curve~\cite{Bala,Schil,Kaneko,Chankova}, which are identified as signatures of a phase transition in a finite system from a strongly paired phase to a phase with weak pairing correlations~\cite{Egido,Liu}. The pairing effect in finite nuclei is well known through the odd-even effect in nuclear masses, for example. On the other hand, in a macroscopic conductor, pairing leads to a phase transition from a normal metal to a superconductor below a certain critical temperature. In the BCS theory~\cite{BCS} of superconductivity the normal to superconducting phase transition is characterized by a finite discontinuity of the heat capacity at the critical temperature. However, in a finite nucleus where the pair coherence length is much larger than the nuclear radius, the sharp discontinuity in heat capacity is smoothed due to large fluctuations. The smoothing of the superfluid to normal phase transition in finite nuclei caused by the non-vanishing of the neutron and/or proton pairing gaps at the critical temperature has been widely discussed in {\it e.g.}, Ref.~\cite{Moretto} and references therein. \\
The Sn isotopes provide an excellent ground to observe the thermal signatures of the pairing interaction. Because of the $Z=50$ shell-closure, the proton pair breaking is strongly hindered, and the signature for the pure neutron pair-breaking becomes much more prominent compared to other mass region~\cite{Agvaan}. In the recent works by the Oslo group, evidences of nucleonic Cooper pair breaking at finite temperates were observed in $^{116}$Sn, $^{117}$Sn, and $^{119}$Sn nuclei~\cite{Agvaan,Toft}. Surprisingly, such features were not prominent for $^{118}$Sn, which could be due to the poorer statistics in this case~\cite{Toft}. It would be interesting to extend similar studies for relatively neutron-deficient Sn isotopes such as $^{115}$Sn.\\
The nucleus $^{115}$Sn is also significant from the astrophysical point of view. It is known to be formed mostly by the astrophysical $p$-process, however, some small contribution from the $s$-process also could not be ruled out~\cite{Arnould}. Presently, the nucleosynthesis calculations largely underpredict the observed abundance of $^{115}$Sn~\cite{Rapp}. Although the origin of the large underproduction could not be attributed to uncertainties in nuclear physics inputs alone, they are seen to influence the nucleosynthesis predictions in a non-negligible way~\cite{Arnould}. The experimental information on critical quantities such as NLD in a wide excitation energy range could be useful to reduce the uncertainty in abundance calculations from the nuclear physics side. \\
\indent In this paper, we report the experimental level density of $^{115}$Sn in the excitation energy range of $\sim $2~-~9 MeV, measured using the neutron evaporation spectra from the $p$~+~$^{115}$In reaction at two compound nuclear excitation energies $E^\ast =$ 18.2 and 21.2 MeV. At these low excitation energies, the compound nucleus ($^{116}$Sn) predominantly decays by the first-chance (1$n$) neutron emission leading to the residual nucleus $^{115}$Sn. Different microscopic models of level density have been tested by comparing with the experimental data. The thermodynamic properties of $^{115}$Sn have also been investigated.\\
The article has been arranged in the following manner. The experimental arrangement has been described in Sec.~\ref{sec2:expt}. The results have been presented and discussed in Sec.~\ref{sec3:result}; the experimental level density and thermodynamic quantities have been presented in Sec.~\ref{sec4:NLD} and Sec.~\ref{sec5:thermo}, respectively. The microscopic EP+IPM calculation has been briefly described in Sec.~\ref{sec6:calc}. Finally, the summary and conclusion have been presented in Sec.~\ref{sec7:Sum}.

\section{\label{sec2:expt} Experimental details}
The experiment was performed using proton beams of $E_\textrm{lab}$~= 9 and 12 MeV from the K130 cyclotron at VECC, Kolkata. A self-supporting foil of $^{115}$In (thickness~$\approx $ 1 mg/cm$^2$) was used as the target. The neutrons emitted during the compound nuclear decay process were detected using six liquid scintillator based neutron detectors (size: 5-inch $\times $~5-inch) placed at the laboratory angles of 55$^\circ $, 85$^\circ $, 105$^\circ $, 120$^\circ $, 140$^\circ $, 155$^\circ $ at a distance of 1.5 m from the target. The neutron kinetic energies were measured using the time-of-flight (TOF) technique. The start trigger for the TOF measurement was generated by detecting the low energy $\gamma $-rays in a 50-element BaF$_2$ detector array~\cite{Deepak} placed near the target position. In converting the neutron TOF to neutron energy, the prompt $\gamma$ peak in TOF spectrum was used as the time reference. The neutron and $\gamma$ separation was achieved by using both the TOF and pulse shape discrimination (PSD) methods. The excitation energy-dependent efficiency, which is a crucial parameter, was measured in the in-beam condition using a standard $^{252}$Cf neutron source~\cite{Pratap3}. The scattered neutron contribution in the measured spectra was estimated using the shadow bar technique and found to be around 2.5$\%$ in the total neutron energy range. It should be mentioned that the back-angle neutron spectra for the present system have been recently utilized to investigate the iso-spin dependence of the level density parameter for $A=$~115 isobars~\cite{PRiso}. Additional information on the experimental setup and analysis techniques can be found in Ref.~\cite{PRiso}. \\
\begin{figure}
\centering
\includegraphics[scale=0.85,clip=true]{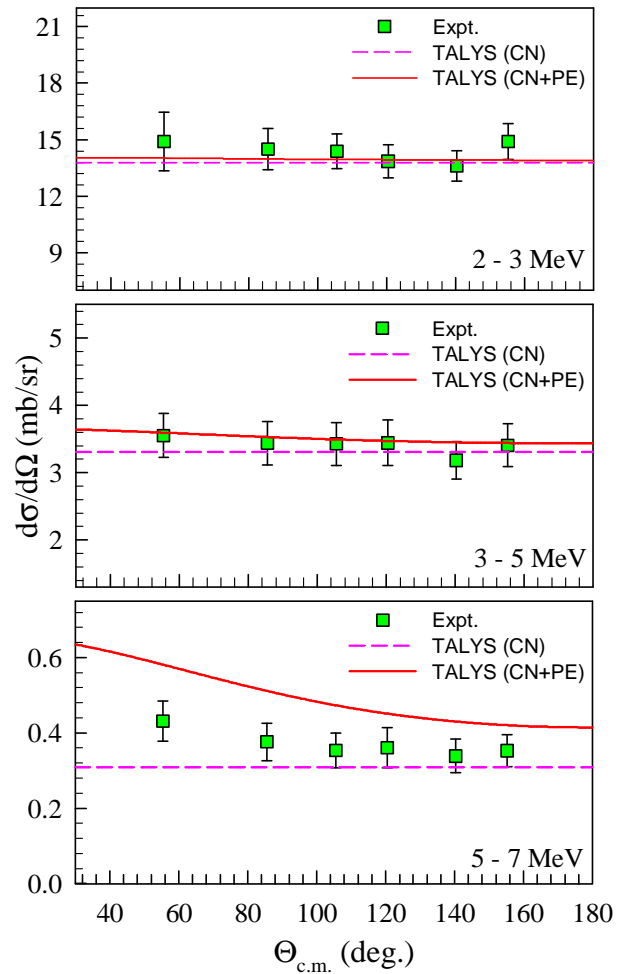}
\caption{\label{fig1:angdistr} (Colour online) Neutron angular distributions for the selected energy bins (indicated in each panel) for the $p$~+~$^{115}$In reaction at $E_\textrm{lab}$~=~12 MeV. The experimental data are shown by the filled squares. The dashed and the continuous lines are the theoretical {\scshape TALYS} prediction for the compound nuclear (CN) and compound plus pre-equilibrium (CN+PE) components, respectively.}
\end{figure}

\section{\label{sec3:result} Results and discussions}
The background-corrected neutron spectra measured at various laboratory angles were transformed into the compound nucleus (CN) center-of-mass (c.m.) frame using the standard Jacobian transformation. The angular distributions in the c.m. frame at the highest bombarding energy ($E_\textrm{lab}$~=12 MeV) for selected neutron energy intervals have been shown in Fig.~1. It can be seen that the cross-section does not vary significantly as a function of the emission angle indicating the dominance of the compound nuclear emission process in determining the neutron spectra at the present incident energies. To investigate the emission mechanisms further, the experimental angular distributions were compared with the theoretical predictions obtained using the {\scshape TALYS} (v~1.9) computer code~\cite{TALYS}. The theoretical TALYS results are shown for both the compound nuclear (dashed lines in Fig.~1) and compound plus pre-equilibrium (PE) (continuous lines in Fig.~1) components. The compound nuclear calculation are performed using the Hauser-Feshbach framework~\cite{Hauser} whereas the PE component has been estimated using the Exciton model~\cite{Preq1,Preq2} in TALYS. It can be seen from Fig.~1 that the contributions of pre-equilibrium processes in the experimental data are small, and it is negligible particularly at the backward angles where the data can be rather described by the CN component alone. The experimental data at the most backward angle (155$^\circ $) is considered almost free of the non-equilibrium component, and used for the statistical model analysis to extract the level density. \\
The spectra measured at 155$^\circ $ for the two incident proton energies of 9 and 12 MeV are shown in Fig.~2. The double-differential cross-sections are multiplied by 4$\pi$ to compare with the calculated angle-integrated spectra. The experimental spectra have been compared with the theoretical {\scshape TALYS} calculations performed within the statistical {\scshape HF} framework. For the level density, the composite Gilbert-Cameron (GC)~\cite{GC} formulation has been used as input of the {\scshape TALYS} calculations. \\
In the GC model, the level density at low energies from 0 up to some matching energy $E_M$ is approximated by a constant-temperature (CT) formula     
\begin{equation}
\rho_{CT}(E )= \frac{1}{T_0}{\rm exp}{\frac{E -E_0}{T_0}}
\end{equation}
and for energies higher than $E_M$ the level density is given by the Fermi gas (FG) expression~\cite{Bethe}, 
\begin{equation}
\rho_{FG} (E )= \frac {1}{12\sqrt{2}\sigma }\frac{\exp{(2\sqrt{a(E -\Delta )})}}{a^{1/4}(E -\Delta )^{5/4}}
\end{equation}
where $a$ is the level density parameter, $\sigma $ is the spin cut-off factor, and $\Delta $ is the pairing energy shift. The constant temperature ($T_0$), and the energy shift ($E_0$) in Eq.~(1) are chosen in such a way the two prescriptions (CT and FG) match together smoothly at the matching energy, which is $\sim $ 4 MeV in the present case~\cite{TALYS}.\\
The shell effect in NLD has been incorporated using an energy and shell-correction dependent parametrization of the level density parameter $a$~\cite{Igna}
\begin{equation}
a(U)=\tilde{a}[1+\frac{\Delta S}{U}\{1-\exp(-\gamma U)\}]
\end{equation}
where $U=E-\Delta $, and $\tilde{a}$ is the asymptotic value of the level density parameter obtained in the absence of any shell effect. Here $\Delta S$ is the ground state shell correction, and $\gamma$ determines the rate at which the shell effect is depleted with the increase in excitation energy. The asymptotic level density parameter $\tilde{a}$ has been tuned to obtain the best match to the experimental data, whereas the energy-shift $\Delta $ was taken from the systematics~\cite{TALYS}. The optimum values of the level density parameter along with other model parameters for the CN ($^{116}$Sn) and major evaporation residues $^{115}$Sn (1$n$) and $^{114}$Sn (2$n$) have been tabulated in Table~\ref{tab:table1}. The contributions of the different decay channels (1$n$ and 2$n$) in the total (inclusive) neutron spectra have been shown in Fig.~2. It can be seen from the figure that at $E^\ast $~=~18.2 MeV ($E_\textrm{lab}$~=~9 MeV) the spectrum is entirely determined by the 1$n$ emission channel whereas at $E^\ast $~=~21.2 MeV ($E_\textrm{lab}$~=~12 MeV) a small contribution from the 2$n$ channel bellow E$_{c.m.}\approx $~3 MeV is observed. \\

\begin{figure}
\centering
\includegraphics[scale=0.75,clip=true]{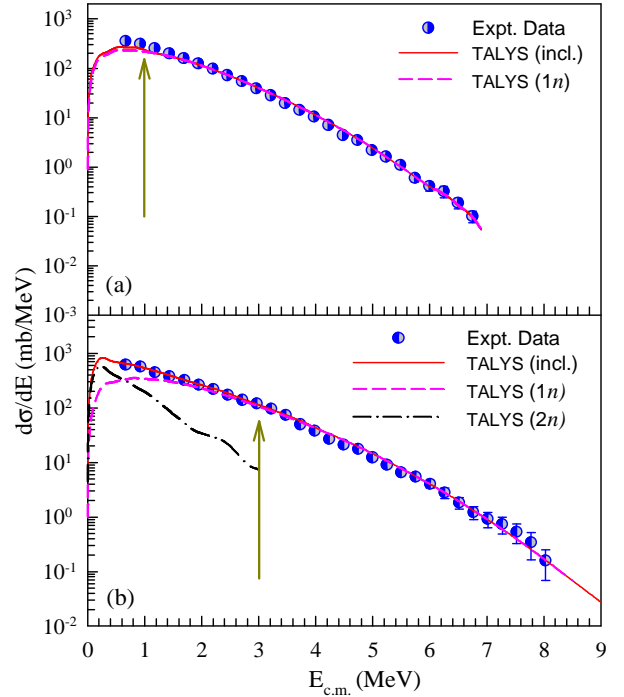}
\caption{\label{fig2:Spectra} (Color online) The experimental neutron energy spectra (symbols) along with the {\scshape TALYS} calculations (lines) at two incident proton energies (a) 9 and (b) 12 MeV. The arrows show the position above which the spectra are entirely determined by the first-chance emission. Contributions from different stages of the decay have also been shown.}
\end{figure}

\begin{center}
\begin{table*}[ht]
{\centering
\caption{\label{tab:table1} Different model parameters used in the {\scshape TALYS} calculation.} {\small \hfill{}
\centering
\begin{tabular}{c c c c c c c c c c c}

\hline
Nucleus &S$_n$& $\tilde{a}$ & $\Delta P$ & $E_0$ &$T_0$ & $E_M$ &$\Delta S$ &$\gamma $ \\
 &(MeV)&(MeV$^{-1}$) &(MeV) & (MeV) &  (MeV)& (MeV)&(MeV)&(MeV$^{-1}$) \\
\hline
\\
$^{116}$Sn &9.563& 15.08 & 2.228 &  1.436 & 0.538 & 5.04 & 1.078 & 0.089 \\
 
\hline
\\
$^{115}$Sn &7.546& 14.95 & 1.119 &  0.566 & 0.511 & 3.49 & 1.008 & 0.089  \\
\hline
\\
$^{114}$Sn &10.303 & 14.82 & 2.248 &  1.116 & 0.599 & 3.49 & 0.682 & 0.089  \\
\hline
\end{tabular}}
\hfill{}
}
\end{table*}
\end{center}

\subsection{\label{sec4:NLD} Nuclear level density of $^{115}$Sn}
The ``experimental" level densities $\rho_{exp}(E)$ were extracted from the neutron spectra using the following relation~\cite{Vonach,Vonach2} 
\begin{equation}
\rho_{exp}(E)= \rho_{model} (E)\frac{(d\sigma /dE)_{exp}}{(d\sigma /dE)_{model}}
\end{equation}
Here $(d\sigma /dE)_{exp}$ is the experimental differential cross-section and $(d\sigma /dE)_{model}$ is the differential cross-section calculated by the {\scshape HF} calculation using $\rho (E)_{model}$ as its input level density. Eq.~4 predicts the correct energy dependence (slope) of level density, and the local variations in NLD are provided by the bin wise normalization factor given by the ratio of the calculated to the experimental differential cross-section. However, the level density can not be obtained in absolute terms by Eq.~4 alone because the ratio $\frac{d\sigma /dE)_{exp}}{(d\sigma /dE)_{model}}$ depends not only on the ratio of $\rho (E)_{exp}$, and $\rho (E)_{model}$ but also on the competition with other decay channels. The calculated cross sections depend on the ratio of the decay probabilities to various channels and an overall arbitrary factor multiplying the NLD for all involved nuclei cancels out in this ratio. Thus one can extract only the energy dependence of the NLD and not their absolute values from the statistical model analysis alone. Therefore, the experimental data obtained using Eq.~4 needs to be renormalized to some known NLD value obtained using other independent techniques. Usually the normalization is done with the density of discrete energy levels for a given nucleus at low energies~\cite{Voinov,Ramirez}. The number of low lying discrete levels provide a less ambiguous and model independent way of normalizing the NLD. However, in the present work, we could not go down to the lowest energies with sufficient statistics. So, the absolute normalization was done at the neutron separation energy ($S_n$) using the level density ($\rho_0$) obtained from the measured $S$-wave neutron resonance spacing ($D_0$). The level density at $S_n$ is related to $D_0$ through the following relation~\cite{Toft}
\begin{equation} 
\begin{split}
\rho _0(S_n)=\frac{2\sigma^2}{D_0}\times [(I_t+1){\rm exp}{\frac{-(I_t+1)^2}{2\sigma^2 }}\\
+I_t {\rm exp}\frac{-I_t^2}{2\sigma^2 }]^{-1}
\end{split}
\end{equation}
\begin{figure}
\centering
\includegraphics[scale= 0.90]{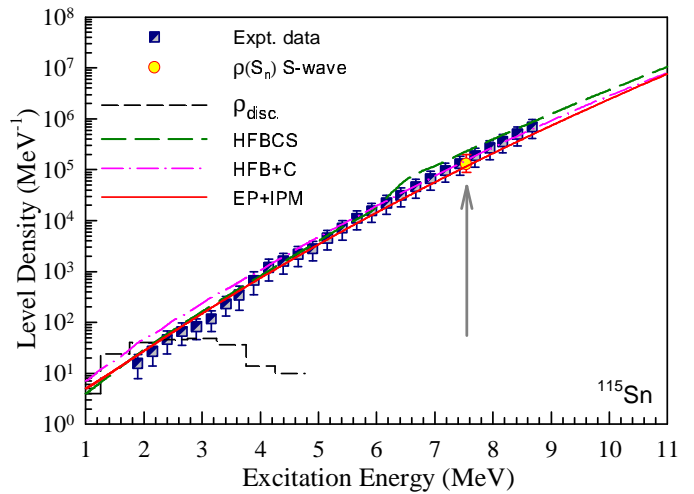}
\caption{(Color online) The experimental level density (filled squares) along with the theoretical predictions (lines) of different microscopic NLD models (see text). The normalization point has been indicated by the arrow.}
\label{fig:fig3}
\end{figure}
Here, $I_t$ is the spin of the target nucleus, and the spin cut-off parameter $\sigma $ is evaluated at $S_n$ using the energy-dependent parametrization prescribed by Egidy and Bucurescu~\cite{Egidy3}. The experimental $S$-wave resonance spacing for $^{115}$Sn is 0.286 ($\pm $~0.110) keV~\cite{RIPL} which gives the normalization point $\rho _0(S_n)=$1.268$\times$10$^5$  MeV$^{-1}$. \\
\indent The experimental level densities for $^{115}$Sn extracted from the neutron spectra have been tabulated in Table~\ref{tab:table2}. The errors in the measured NLD represents both the statistical and systematic uncertainties. A maximum systematic error of 40$\%$ arising out of the uncertainty in the determination of the normalization point has been added in quadrature with the statistical errors. The experimental level densities are also plotted in Fig.~3 by filled squares. The density of discrete levels ($\rho_\textrm{disc.}$) obtained from the experimental energy levels~\cite{RIPL} of $^{115}$Sn is also plotted in Fig.~3 by the short-dashed line. In extracting the level densities using Eq.~4, only the part of the experimental spectra which is determined completely by the first-chance neutron emission (shown by the arrows in Fig.~2) was used. The excitation energy of the residual nucleus ($E$) was calculated using the relation $E=E^*-S_n-E_n-E_R$, where $E^*$ is the compound nuclear excitation energy, $S_n$ is the neutron separation energy, $E_n$ and $E_R$ are the kinetic energies of the emitted neutron and the residual nucleus, respectively. The residual nucleus ($^{115}$Sn) being significantly heavier than the emitted neutron its kinetic energy is negligible. For the overlapping data points obtained from the neutron spectra measured at the two bombarding energies, the average values of the level densities have been taken. \\
\begin{table}[ht]
\centering
\caption{\label{tab:table2} Total nuclear level densities in $^{115}$Sn.} {\small \hfill{}
\centering
\begin{tabular}{c c c }
\hline
Excitation energy ($E$)  & Level density ($\rho (E)$) \\
 (MeV)     &  $\times $10$^3$ (MeV$^{-1}$)    \\

 \hline
\\
1.90	&	0.016	~$\pm$~	0.008\\
2.14	&	0.027	~$\pm$~	0.013\\
2.40	&	0.046	~$\pm$~	0.022\\
2.65	&	0.066	~$\pm$~	0.031\\
2.90	&	0.082	~$\pm$~	0.037\\
3.16	&	0.117	~$\pm$~	0.050\\
3.41	&	0.229	~$\pm$~	0.096\\
3.64	&	0.342	~$\pm$~	0.168\\
3.89	&	0.671	~$\pm$~	0.321\\
4.14	&	1.196	~$\pm$~	0.536\\
4.39	&	1.592	~$\pm$~	0.694\\
4.65	&	2.176	~$\pm$~	0.937\\
4.90	&	2.779	~$\pm$~	1.173\\
5.15	&	4.509	~$\pm$~	1.885\\
5.40	&	7.067	~$\pm$~	2.930\\
5.66	&	11.05	~$\pm$~	4.510\\
5.91	&	15.53	~$\pm$~	6.297\\
6.16	&	22.15	~$\pm$~	8.978\\
6.42	&	31.33	~$\pm$~	12.63\\
6.67	&	46.47	~$\pm$~	18.74\\
6.92	&	66.43	~$\pm$~	26.70\\
7.18	&	95.85	~$\pm$~	38.48\\
7.43	&	130.7	~$\pm$~	52.50\\
7.69	&	186.6	~$\pm$~	74.86\\
7.94	&	265.8	~$\pm$~	106.5\\
8.16	&	357.6	~$\pm$~	143.8\\
8.42	&	496.7	~$\pm$~	199.4\\
8.67	&	688.5	~$\pm$~	276.2\\
\hline
\end{tabular}}
\hfill{}
\end{table}
The experimental level densities provide an excellent opportunity to test different microscopic approaches of NLD. The Hartree-Fock-BCS (HFBCS)~\cite{Goriely1,Demet}, and Hartree-Fock-Bogoliubov plus combinatorial (HFB+C)~\cite{Goriely2} models are the most frequently used microscopic methods that provide a global description of NLD. The experimental data have been compared with the predictions of HFBCS and HFB+C methods, as well as an exact pairing plus independent particle model (EP+IPM), introduced recently~\cite{Dang,Dang2}. It is observed that the experimental data are in excellent agreement with the prediction of the microscopic EP+IPM (shown by the (red) continuous line in Fig.~3) which takes into account the thermal effects of exact pairing in finite size systems~\cite{Dang}. The HFB+C calculation of Goriely {\it et al.},~\cite{Goriely2} also explains the experimental data reasonably well except for a small disagreement at lower energies as shown by the dash-dotted line in Fig.~3. On the other hand, the total level densities calculated based on the single-particle level schemes determined within the HFBCS model~\cite{Goriely1,Demet} deviates from the experimental data at higher energies (medium-dashed line in Fig.~3). The agreement of EP+IPM with the experimental data emphasizes the crucial role of thermal pairing in the description of the NLD. It should be mentioned here, that unlike the HFBCS and HFB+C methods, the EP+IPM-predicted NLDs do not require any additional normalization while comparing with experimental data.\\

\subsection{\label{sec5:thermo} Thermal properties of $^{115}$Sn}
The thermal properties of $^{115}$Sn were investigated using the measured level densities. The microcanonical entropy ($S$), and temperature ($T$) were extracted from the measured NLDs following the prescription of several earlier works~\cite{Melby2,Giacoppo,Toft}. The entropy is defined by
\begin{equation}
S = k_B {\rm ln} \left[\frac{\rho (E)}{\rho_0}\right]
\end{equation}
where $k_B$ is the Boltzmann constant which is usually set to unity to make the entropy dimensionless. The constant $\rho _0$ is adjusted to fulfill the condition of the third law of thermodynamics $S\rightarrow  $0 as $T\rightarrow  $0. The ground states of even-even nuclei represent a completely ordered system with only one possible configuration, and are characterized by zero entropy and temperature. The value of $\rho _0=$~0.135 MeV$^{-1}$ thus obtained for the nearest even-even nuclei $^{116}$Sn~\cite{Agvaan} is also used in the present case. The microcanonical temperature ($T$) is defined by the relation
\begin{figure}
\centering
\includegraphics[scale= 0.75]{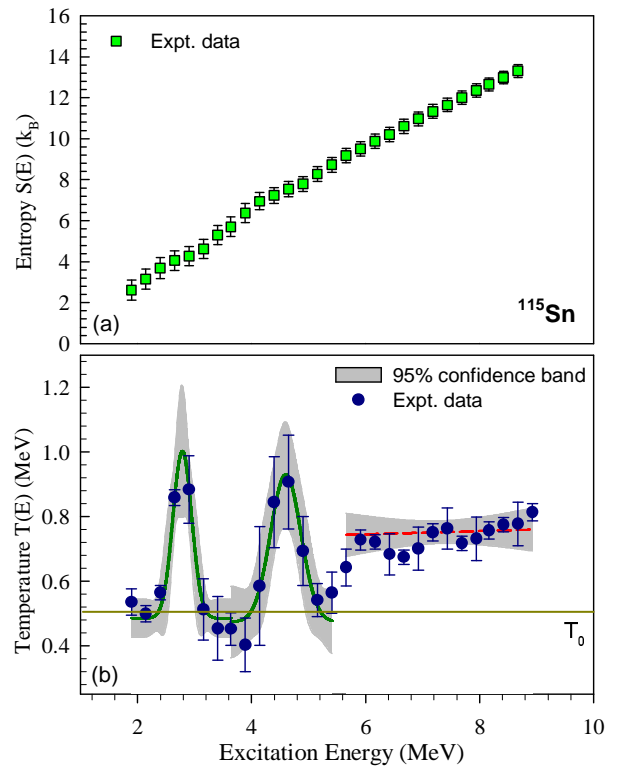}
\caption{(Color online) The experimental (a) entropy and (b) temperature as a function of excitation energy. The solid (green) and the dashed (red) lines are the Gaussian and linear fits to the data, respectively. The gray shaded regions represent the 95$\%$ confidence band. }
\label{fig:fig4}
\end{figure}
\begin{equation}
T = \left(\frac{\partial S}{\partial E}\right)^{-1}
\end{equation}
The experimental entropy and temperature as a function of the excitation energy have been plotted in Fig.~4. The number of accessible microstates increases as a function of the excitation energy resulting in the gross increase of entropy as shown in Fig.~4(a). Any possible fine-structure present in the entropy curve will be amplified in the temperature plot due to the derivative relation with entropy. The temperature profile, thus obtained is displayed in Fig.~4(b), which shows fluctuating nature characterizing a small isolated system like the atomic nucleus, away from thermodynamic limits. On top of the inherent fluctuations, there are a couple of distinct peak-like structures in the temperature profile between $E\approx  $2~-~6 MeV. Further, the temperature seems to remain almost constant in the region beyond $E\sim $6 MeV as indicated by the dashed line in Fig.~4(b). Here, the first and second peaks in the microcanonical temperature profile, which are centered around E $\sim $~2.4 and 4.4 MeV, are possibly associated to the breaking of the second and third Cooper pairs, similar to those observed previously in $^{116,117}$Sn~\cite{Agvaan}.\\ 

\subsection{\label{sec6:calc} Microscopic calculations}
To investigate the microscopic origin of the observed features in the temperature plot, we have calculated the temperature dependent neutron pairing gap ($\Delta P_n$) within the EP+IPM which provided the best description of the experimental level density (Fig.~3). The EP+IPM formalism for the calculations of different thermodynamic quantities has been described in detail in Refs.~\cite{Bala,Dang,Dang2}. The calculated pairing gap, as shown in Fig.~5(a), decreases rapidly in the region between $E \sim  $1 to 6 MeV (as shown in the inset of Fig.~5(a)) around the critical temperature $T_c$ ($ \sim $0.6 MeV) which is defined as the temperature at which the pairing gap vanishes according to the conventional BCS theory~\cite{BCS}. This is the region of the pairing phase transition in which the first few Cooper pairs are broken resulting in the peak-like structures in the temperature curve. The enhancement of the pairing around $T =$~0.4 MeV is due to the blocking effect at finite temperature in odd nucleon systems~\cite{Hung}. When more Cooper pairs are broken, the pairing gap decreases slowly, which is reflected in the near constant nature of the experimental temperature in the region beyond $E\approx $ 6 MeV (Fig.~4(b)). The pairing gap does not disappear completely due to the thermal fluctuations in finite size systems~\cite{Dang}.  \\
\begin{figure}
\centering
\includegraphics[scale= 0.70]{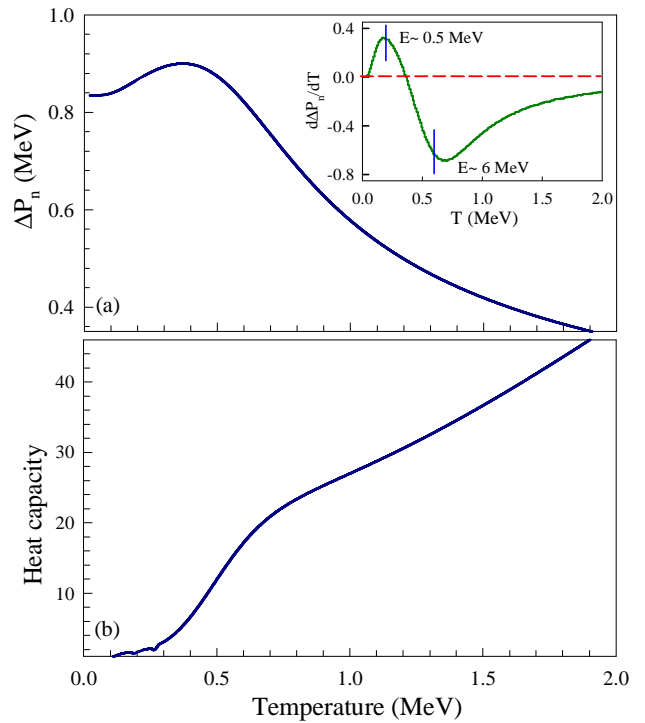}
\caption{(Color online) (a) The neutron pairing gap, and (b) the canonical heat capacity ($C_v$) of $^{115}$Sn as a function of temperature obtained within the EP+IPM calculation. The rate of change of the neutron pairing gap as a function of the temperature is shown in the inset of the upper panel.}
\label{fig:fig5}
\end{figure}
The effect of the sequential neutron pair breaking in NLD and the thermodynamic quantities for $^{115}$Sn has been investigated within the EP+IPM, and it is found that the different regions of the excitation energy can be identified as, (i) $E\sim $~0~-~0.5 MeV: blocking effect of odd neutron, (ii) $E\sim $~0.5~-~2 MeV: breaking of one neutron pair, (iii) $E\sim $~2~-~4 MeV: breaking of two neutron pairs, and (iv) $E\sim $~4~-~6 MeV: breaking of three neutron pairs. Hence, the peaks in the temperature profile as shown in the Fig.~4(b) are caused by the breaking second and third neutron pairs in $^{115}$Sn. In general, for odd mass nuclei ($e.g.$ $^{115}$Sn and $^{117}$Sn~\cite{Agvaan}) there exist two peaks in the microcanonical temperature curve below E $<$~2 MeV. The first peak is related to the blocking of the odd neutron (or the very low-lying excited states), whereas the second peak is associated with the breaking of the first pair. As the present experimental data starts from 2 MeV the peak due to the breaking of the first neutron pair could not be observed. Hence, the two peaks around E$\sim $~2~-~6 MeV observed in Fig. 4(b) should correspond to the breaking of the second and third neutron pairs.\\
\indent Along with the neutron pairing gap the canonical heat capacity has also been calculated within the EP+IPM, and shown in Fig.~5(b). The calculated heat capacity shows a S-shape nature with a local maximum around the critical temperature. As described in recent theoretical works~\cite{Egido,Liu}, the S-shaped heat capacity curve could be treated as a fingerprint of a phase transition in a finite system from a phase with strong pairing correlations to a phase with weak pairing correlations. Pronounced S-shapes in heat capacity are generally expected for the even-even nuclei, whereas for neighboring odd-odd and odd-even nuclei such structures could be less prominent~\cite{Melby2,Bala,Chankova}. In the present case of the even-odd $^{115}$Sn, there is a distinct enhancement in the associated heat-capacity in around $T\approx $~0.6 MeV providing a clear signature of the pairing phase transition in $^{115}$Sn.  
\\
\section{\label{sec7:Sum} Summary and conclusions}
The neutron energy spectra in case of $p$~+~$^{115}$In reaction have been measured at low incident energies, and compared with the statistical Hauser-Feshbach calculation performed using the {\scshape TALYS} code. The nuclear level density of the residual $^{115}$Sn nucleus has been extracted in the excitation energy range of $\sim $2~-~9 MeV from the back-angle neutron spectra. The experimental level density has been compared with the predictions of microscopic HFBCS, HFB+C, and EP+IPM models. It is observed that the experimental data are best described by the EP+IPM calculation highlighting the importance of thermal pairing in the description of NLD. The experimental temperature profile shows peak-like structures around $E\sim $2~-~6 MeV which are linked with breaking of nucleonic Cooper pairs. The calculated heat capacity for $^{115}$Sn showed a S-shape which provide a clear indication of the quenching of the nuclear pairing correlation at finite temperature. \\  
\section*{Acknowledgments}
The authors would like to acknowledge the VECC Cyclotron operators for smooth running of the accelerator during the experiment. We are thankful to J. K. Meena, A. K. Saha, J. K. Sahoo and R. M. Saha for their help during the experimental setup. The authors also thanks Jhilam Sadhukhan for the stimulating discussions.\\  
NQH's works are funded by The National Foundation for Science and Technology Development of Vietnam through Grant Number 103.04-2019.371.


\normalsize
\begin{thebibliography}{99}
\bibitem{Hauser} W. Hauser and H. Feshbach, Phys. Rev. {\bf 87}, 366 (1952).
\bibitem{Cecilie} A.C. Larsen, A. Spyrou, S.N. Liddick, M. Guttormsen, Progress in Particle and Nuclear Physics {\bf 107} (2019) 69-108.
\bibitem{Arnould} M. Arnould, K. Takahashi, Rep. Prog. Phys. {\bf 62} (1999) 395.
\bibitem{Bowman} C. D. Bowman, Annu. Rev. Nucl. Part. Sci. {\bf 48}, 505 (1998).
\bibitem{Mirz} H. R. Mirzaei, et al., J. Cancer Res. Ther. {\bf 12}, 520 (2016).
\bibitem{Melby} E. Melby, L. Bergholt, M. Guttormsen, {\it et al.} Phys. Rev. Lett. {\bf 83}, 3150 (1999).
\bibitem{Giacoppo} F. Giacoppo, F. L. Bello Garrote, L. A. Bernstein, Phys. Rev. {\bf C 90}, 054330 (2014).
\bibitem{Melby2} E. Melby, M. Guttormsen, J. Rekstad, Phys. Rev. {\bf C 63}, 044309 (2001).
\bibitem{Agvaan} U. Agvaanluvsan, A. C. Larsen, M. Guttormsen {\it et al.} Phys. Rev. {\bf C 79}, 014320 (2009).
\bibitem{Toft} H. K. Toft, A. C. Larsen, U. Agvaanluvsan, {\it et al.} Phys. Rev. {\bf C 81}, 064311 (2010).
\bibitem{Syed} N. U. H. Syed, A. C. Larsen, A. Burger {\it et al.}, Phys. Rev. {\bf C 80}, 044309 (2009).
\bibitem{Gutt} M. Guttormsen, B. Jurado, J. N. Wilson {\it et al.}, Phys. Rev. {\bf C 88}, 024307 (2013).
\bibitem{Bala} Balaram Dey, N. Quang Hung, Deepak Pandit {\it et al.}, Phys. Lett. {\bf B 789} (2019) 634-638.
\bibitem{Schil} A. Schiller, A. Bjerve, M. Guttormsen {\it et al.} Phys. Rev. {\bf C 63}, 021306 (2001).
\bibitem{Kaneko} K. Kaneko, M. Hasegawa, U. Agvaanluvsan, Phys. Rev. {\bf C 74}, 024325 (2006).
\bibitem{Chankova} R. Chankova, A. Schiller, U. Agvaanluvsan, Phys. Rev. {\bf C 73}, 034311 (2006).
\bibitem{Egido} J. L. Egido, L. M. Robledo, and V. Martin, Phys. Rev. Lett. {\bf 85}, 26 (2000).
\bibitem{Liu} S. Liu and Y. Alhassid, Phys. Rev. Lett. {\bf 87}, 022501 (2001).
\bibitem{BCS} J. Bardeen, L.N. Cooper, J.R. Schrieffer, Phys. Rev. {\bf 108} (1957) 1175.
\bibitem{Moretto} N. Quang Hung, N. Dinh Dang, and L. G. Moretto, Rep. Prog. Phys. {\bf 82}, 056301 (2019) . 
\bibitem{Rapp} W. Rapp, J. Gorres, M. Wiescher, {\it et al.} Astrophys. J. {\bf 653} (2006) 474.
\bibitem{Deepak} Deepak Pandit, S. Mukhopadhyay, Srijit Bhattacharya {\it et al.}, Nucl. Instrum. Methods Phys. Res. {\bf A 624}, 148 (2010).
\bibitem{Pratap3} Pratap Roy, K. Banerjee, A. K. Saha {\it et al.} Nuclear Inst. and Methods in Physics Research, {\bf A 901}, (2018) 198-202.
\bibitem{Sujoy} S. Chatterjeea, K. Banerjeeb, Deepak Pandit {\it et al.}, Applied Radiation and Isotopes, {\bf 128} (2017) 216-223.
\bibitem{PRiso} Pratap Roy, K. Banerjee, T. K. Rana {\it et al.}, Phys. Rev. {\bf C} (Accepted, in production).
\bibitem{TALYS} A.J. Koning, S. Hilaire, and M.C. Duijvestijn, {\bf TALYS-1.9}, online at $www.talys .eu.$
\bibitem{Preq1} A.J. Koning and M.C. Duijvestijn, Nucl. Phys. {\bf A 744}, 15 (2004).
\bibitem{Preq2} E. Gadioli and P.E. Hodgson, Pre-equilibrium nuclear reactions, Oxford Univ. Press (1992).
\bibitem{GC} A. Gilbert and A.G.W. Cameron, Can. J. Phys. {\bf 43}, 1446 (1965).
\bibitem{Bethe} H. A. Bethe, Phys. Rev. {\bf 50}, 332 (1936); Rev. Mod. Phys. {\bf 9}, 69 (1937).
\bibitem{Igna} A.V. Ignatyuk, G.N. Smirenkin and A.S. Tishin, Sov. J. Nucl. Phys. {\bf 21}, 255 (1975).
\bibitem{Vonach} H. Vonach, Proceedings of the IAEA Advisory Group Meeting on Basic and Applied Problems of Nuclear Level Densities, Upton, New York, 1983, Brookhaven National Laboratory Report No. {\bf BNL-NCS-51694}, 1983, p. 247.
\bibitem{Vonach2} A. Wallner, B. Strohmaier, and H. Vonach, Phys. Rev. {\bf C 51}, 614 (1995).
\bibitem{Voinov} A. V. Voinov, S. M. Grimes, C. R. Brune {\it et al.}, Phys. Rev. {\bf C 76}, 044602 (2007).
\bibitem{Ramirez} A. P. D. Ramirez, A. V. Voinov, S. M. Grimes {\it et al.}, Phys. Rev. {\bf C 88}, 064324 (2013).
\bibitem{Egidy3} Till Von Egidy and Dorel Bucurescu, Phys. Rev. {\bf C 80}, 054310 (2009).
\bibitem{RIPL} R. Capote, M. Herman, P. Oblozinsky, {\it et al.}, Nuclear Data Sheets, {\bf 110}, 3107 (2009).
\bibitem{Goriely1} S. Goriely, F. Tondeur, J.M. Pearson, Atomic Data Nuclear Data Tables {\bf 77}, 311 (2001).
\bibitem{Demet} P. Demetriou, S. Goriely, Nucl. Phys. {\bf A 695}, 95 (2001).
\bibitem{Goriely2} S. Goriely, S. Hilaire, and A.J. Koning, Phys. Rev. {\bf C 78} (2008) 064307.
\bibitem{Dang} N. Quang Hung, N. Dinh Dang, and L. T. Quynh Huong, Phys. Rev. Lett. {\bf 118}, 022502 (2017).
\bibitem{Dang2} N. Dinh Dang, N. Quang Hung, and L. T. Quynh Huong, Phys. Rev. {\bf C 96}, 054321 (2017).
\bibitem{Hung} N. Quang Hung, N. Dinh Dang, and L. T. Quynh Huong, Phys. Rev. {\bf C 94}, 024341 (2016).
\end{thebibliography}
\end{document}